\pdfoutput=1
\documentclass[twocolumn,
               showpacs,
               preprintnumbers,
               nofootinbib,
               prd,
               superscriptaddress,
               10pt,
               notitlepage,
               aps]{revtex4-2}
               
\usepackage{graphicx,amssymb,amsmath,amsthm,amsfonts,epsfig,mathtools}

\usepackage[utf8]{inputenc}
\usepackage{graphicx}
\usepackage{dcolumn}
\usepackage{bm}
\usepackage{amsmath}
\usepackage{color}
\usepackage[dvipsnames]{xcolor}
\usepackage{hyperref}
\hypersetup{colorlinks=true, citecolor=MidnightBlue,linkcolor=CornflowerBlue, urlcolor=CornflowerBlue, linktocpage=true}
\usepackage{xfrac}
\usepackage{siunitx}
\usepackage{soul}
\newcommand{\beqn}{\begin{eqnarray}}
\newcommand{\eeqn}{\end{eqnarray}}
\newcommand{\beq}{\begin{equation}}
\newcommand{\eeq}{\end{equation}}

\def\gbar{\bar{g}}
\def\gnn{\bar{g}_{nn}}
\def\gnn{\bar{g}_{nn}}

\def\cala{{\cal A}}

\begin{document}

\title{Coordinate singularities of self-interacting vector field theories}

\begin{abstract}
Self-interacting vectors are seeing a burst of interest where various groups demonstrated that the field evolution ends in finite time. Two nonequivalent criteria have been offered to identify this breakdown: (i) the vector constraint equation cannot be satisfied beyond a point where the breakdown occurs, (ii) the dynamics is governed by an effective metric that becomes singular at the breakdown. We show that (i) identifies a coordinate singularity, and can be removed by a change of coordinates. Hence, it does not signify a physical problem, and cannot determine the validity of a theory.

\end{abstract}

\author{Andrew Coates}
\email{acoates@ku.edu.tr}
\affiliation{Department of Physics, Ko\c{c} University, \\
Rumelifeneri Yolu, 34450 Sariyer, Istanbul, Turkey}

\author{Fethi M. Ramazano\u{g}lu}
\email{framazanoglu@ku.edu.tr}
\affiliation{Department of Physics, Ko\c{c} University, \\
Rumelifeneri Yolu, 34450 Sariyer, Istanbul, Turkey}

\date{\today}
\maketitle

\noindent
Self-interacting vector fields find applications in many areas of physics. One can find them in gravity and cosmology~\cite{Esposito-Farese:2009wbc,DeFelice:2016yws,DeFelice:2016cri,DeFelice:2016uil,Emami:2016ldl,Heisenberg:2017hwb,Kase:2017egk,Ramazanoglu:2017xbl,Annulli:2019fzq,Barton:2021wfj,Minamitsuji:2018kof, Herdeiro:2020jzx,Herdeiro:2021lwl,Garcia-Saenz:2021uyv,Silva:2021jya,Demirboga:2021nrc,Doneva:2022ewd}, plasma physics~\cite{RevModPhys.78.591}, astrophysics~\cite{Conlon:2017hhi,Fukuda:2019ewf,dEnterria:2013zqi, Burgess:2020tbq} and effective models of photon-photon interactions~\cite{Heisenberg:1936nmg, ATLAS:2017fur,RevModPhys.78.591}. Particularly in the gravity and cosmology communities there have been attempts to classify all ghost-free generalizations of the Proca theory~\cite{Proca:1936fbw,Heisenberg:2014rta, Heisenberg:2016eld, Kimura:2016rzw, Allys:2015sht}.

However, there is a recently growing literature demonstrating that non-linear vector field theories, even the simplest conceivable extensions of the Proca theory, suffer from fatal problems in terms of time evolution~\cite{Esposito-Farese:2009wbc,Clough:2022ygm,Mou:2022hqb,Coates:2022qia}. Most recently, it was shown that these problems can appear in an unusual way, where the self-interacting vectors can evolve without any issue for a finite time, but the time evolution breaks down when the field amplitude reaches certain finite values~\cite{Clough:2022ygm,Mou:2022hqb,Coates:2022qia}. This offers new purely theoretical tests of field theories, which can help guide theory building in the aforementioned research areas. Despite these exciting developments, there are still major points of confusion in the literature, which we aim to address in this letter.

The breakdown of the time evolution we mentioned has been identified by two separate methods in the most recent studies. In the first, which we will call the \emph{constraint criterion}, one uses the fact that the time evolution of the vector is a constrained one, and there comes a point where the satisfaction of the constraint becomes impossible, which is interpreted as the breakdown of time evolution~\cite{Clough:2022ygm,Mou:2022hqb}. In the second, which we will call the \emph{metric singularity criterion}, one shows that the dynamics of the vector is governed by an \emph{effective metric} which can become singular due to its dependence on the vector field itself, and time evolution is not possible beyond such a singular point~\cite{Esposito-Farese:2009wbc,Coates:2022qia}. 

In the following, we will explicitly demonstrate that the constraint criterion indicates a coordinate singularity which does not point to a physical problem, i.e. the theory can evolve beyond such a point if appropriate coordinates are utilized. Hence, this approach does not indicate a physical breakdown of self-interacting vector field theories, or other theories for which similar criteria exist.

The constraint criterion might seem to be the more natural one if one uses the common technique of $3+1$ decomposition for time evolution, and it can even seem to be the \emph{only} criterion on the flat Minkowski background where the effective metric and its curvature might be easily overlooked. Hence, the coordinate-dependent nature of the constraint is subtle, and likely contributed to the confusion in the literature.\footnote{Earlier preprints of \textcite{Coates:2022qia} also suffered from this confusion, which was corrected in the later versions and the published manuscript.} As a result, even the fact that there are two separate criteria, let alone they are nonequivalent, has not been appreciated so far. Overall, our findings are crucial in obtaining accurate diagnostics for the problems of self-interacting vectors or other constrained fields.

We use the ``mostly plus" metric signature and $c=1$. Spacetime indices are in Greek, $\mu,\nu=0,1, \dots ,d$, spatial ones in Latin $i,j=1,\dots , d$.

\noindent {\bf \em   Singularities of the nonlinear Proca theory:}
Let us first overview the problems of self-interacting vectors in terms of the two criteria above following \textcite{Coates:2022qia}.

The simple extension of the Proca theory we will study is given by the action
\begin{align}\label{eq:action}
    {\cal L} = -\frac{1}{4} F_{\mu\nu}F^{\mu\nu} - \overbrace{\left( \frac{\mu^2}{2} X^2 + \frac{\lambda \mu^2}{4} \left( X^2\right)^2 \right)}^{V(X^2)} \ ,
\end{align}
with $F_{\mu\nu} = \nabla_\mu X_\nu - \nabla_\nu X_\mu$ and $X^2 =X_\mu X^\mu$ for the real vector field $X_\mu$. The fields live on a fixed curved spacetime with metric $g_{\mu\nu}$, which is the metric that lowers and raises tensor indices, and defines the connection. This leads to the equation of motion
\begin{align}\label{eq:eom}
    \nabla_\mu F^{\mu\nu} = \mu^2 z X^{\nu} \ ,
\end{align}
where $z=2V'/\mu^2=1+\lambda X^2$ and $V'=(dV/dX^2)$. This also implies the (generalized) Lorenz condition
\begin{align}\label{eq:lorenz}
    \nabla_\nu \nabla_\mu F^{\mu\nu} = 0\ 
    \Rightarrow\ \nabla_\mu \left(z X^\mu \right) = 0 
\end{align}
due to the antisymmetry of $F_{\mu\nu}$. Even though we chose a specific form of $V(X^2)$, a generic choice leads to similar conclusions~\cite{Mou:2022hqb,Clough:2022ygm,Coates:2022qia}.

The key observation for the metric singularity criterion is that Eq.~\eqref{eq:eom} can be put into the form~\cite{Coates:2022qia}\footnote{Even though they use the constraint criterion as we will discuss, the effective metric idea was first adapted to the extensions of the Proca theory by \textcite{Clough:2022ygm}, following earlier work on spontaneous vectorization~\cite{Silva:2021jya,Demirboga:2021nrc,Doneva:2022ewd}.}
\begin{align}\label{eq:gbar_eom}
    \gbar_{\alpha\beta}\nabla^\alpha \nabla^\beta X_\nu + \dots 
    = 0 \ 
\end{align}
with the help of the Lorenz condition, where the ellipses are lower order terms in derivatives. In other words,  the principle part of the differential equation is the wave operator for the effective metric
\begin{align}
    \gbar_{\mu\nu} &= z g_{\mu\nu} + 2z' X_\mu X_\nu\ , \label{eq:g_eff} 
\end{align}
which controls the dynamics. Strictly speaking, this is only possible in $1+1$D, however it can be shown that the effective metric still governs the dynamics in any dimension, through other methods~\cite{Coates:2022qia}. 

Eq.~\eqref{eq:gbar_eom} means the behavior of solutions are as if $X_\mu$ is evolving in a spacetime with metric $\gbar_{\mu\nu}$. The effective metric depends on the vector field itself, hence, can become singular or change its signature at finite values of $X_\mu$, and the time evolution can break down in finite duration even if $g_{\mu\nu}$ is regular~\cite{Esposito-Farese:2009wbc,Coates:2022qia}. Hence, the time evolution cannot continue to the future of such a point, the same way it cannot continue to the future of a singularity in the spacetime metric. The singularity of $\gbar_{\mu\nu}$ can be mathematically determined by finding the points with
\begin{equation}\label{eq:detg}
    \gbar = g \left(1+\lambda X^2 \right) \left(1+3\lambda X^2 \right) = g\ z\ z_3 = 0
\end{equation}
where $g = \det(g_{\mu\nu})$, $z_3 = 1+3\lambda X^2$. Thus, hyperbolicity is lost when $z_3=0$, which always occurs before $z=0$ for physically meaningful initial data~\cite{Coates:2022qia}. In general, vanishing of the determinant of a metric can be a coordinate effect, however, it is known that this case corresponds to a curvature singularity in $\gbar_{\mu\nu}$, hence time evolution indeed cannot continue beyond $z_3=0$ in any formulation of the theory~\cite{Coates:2022qia}.

To understand the constraint criterion, we first note that the very concept of time evolution requires choosing a timelike direction on the spacetime manifold. This means picking a specific coordinate system and a \emph{foliation}, in its most common form an expression of the spacetime as a combination of spatial surfaces stacked in the time direction~\cite{Arnowitt:1962hi,gourgoulhon20123+1}. This is commonly used with the so-called $d+1$ decomposition in $d+1$ dimensions, where tensors are also expressed in terms of their temporal and spatial components
\begin{align}\label{eq:adm}
    ds^2 &= -\alpha^2 dt^2 +\gamma_{ij} (dx^i + \beta^i dt)(dx^j + \beta^j dt) \\
    X_\mu &= n_\mu \phi + A_\mu \  ,  \ \phi = -n_\mu X^\mu \  ,  \ A_i=\left(\delta^{\mu}{}_{i} +n^\mu n_i \right)X_\mu \ . \nonumber
\end{align}
where $n^\mu =\alpha^{-1}(1,-\beta^i)$ is the normal vector field to the spatial slices.

After some lengthy but standard algebra, the relevant part of the time evolution equations can be recast as~\cite{Clough:2022ygm}
\begin{align}\label{eq:eom1p1}
    \partial_t \phi &= \beta^i D_i \phi -A^i D_i \alpha - \frac{\alpha}{\gnn} z \left(K\phi -D_i A^i \right) \\
     +\frac{2\lambda \alpha}{\gnn} &\left[A^i A^j D_i A_j -\phi \left(E_i A^i - K_{ij} A^i A^j + 2A^i D_i \phi \right) \right] \nonumber \\
    0&= D_i E^i + \mu^2 z \phi = {\cal C}  \nonumber \ 
\end{align}
where $E_i= \left(\delta^{\mu}{}_{i} +n^\mu n_i \right)n^\nu F_{\mu\nu}$, the first equation is a result of Eq.~\eqref{eq:lorenz}, and the second one is the component of Eq.~\eqref{eq:eom} along $n^\mu$. Note that the last line, $\mathcal{C}=0$, called the \emph{constraint equation}, does not represent time evolution, but is a necessary condition which has to be satisfied by the vector field components on each spatial slice. Further details of this time-space decomposition can be found in standard sources~\cite{gourgoulhon20123+1}, but are not essential for our purposes. 

It is straightforward to note that $\partial_t \phi$ diverges when 
\begin{align}\label{eq:gnn}
    \gnn &= n^\mu n^\nu \gbar_{\mu\nu} = -z_3 +2\lambda A_iA^i = 0\ ,
\end{align}
which means time evolution cannot continue beyond such a point, which is the constraint criterion. Our naming of this criterion is due to the fact that the constraint equation $\mathcal{C}=0$ ceases to have a unique solution exactly when $\gnn=0$ occurs, which can also be interpreted as the underlying reason for the problem in the time evolution of $\phi$~\cite{Mou:2022hqb,Clough:2022ygm}.

Below, we will demonstrate an explicit example of the coordinate-dependent nature of the constraint criterion by considering a wave packet that falls into a black hole. In one coordinate choice, $\gbar_{\tilde{n}\tilde{n}}=0$ is encountered in finite time and the time evolution indeed has to be stopped, but the problem disappears once different coordinates are used. This clearly shows that the problem indicated by the constraint criterion is not about the physical nature of the vector field, but about the shortcomings of the time evolution method, in this case the particular spacetime foliation, one chooses. That is, a point with $\gnn=0$ can be transformed into one with $\gbar_{\tilde{n}\tilde{n}}<0$ with an alternative foliation defined by a new normal vector field $\tilde{n}^\mu$~\cite{Coates:2022qia}. This point was conceptually argued by \textcite{Coates:2022qia}, but no explicit example was known until now.

\noindent {\bf \em Appearance and disappearance of coordinate singularities:}
What happens when a self-interacting vector wave packet falls into a black hole? The equivalence principle would predict that nothing physically dramatic occurs, since nothing dramatic happens in flat space~\cite{Coates:2022qia}, aside from some possible ``tidal'' effects arising from nonzero packet size. We will see that this is indeed the case as far as physics goes, but a thorough demonstration is quite nontrivial since the constraint criterion is eventually satisfied in the generic case if we use the most familiar coordinates. Thus, coordinate singularities appear, but we will also see that they disappear with a careful choice of coordinates.

The coordinate dependent nature of the constraint criterion can be demonstrated on the spacetime of a $1+1$D black hole. The metric we study is simply the $rt$ section of the Schwarzschild metric
\begin{align} 
    ds^2 &= -f(r) dt^2 + dr^2/f(r)  \\
   & = \Omega^2(x) \left(-dt^2 + dx^2 \right) \ , \label{eq:11metric}
\end{align}
where $f(r)=1-\sfrac{2M}{r}$. In the following, we will use the conformally flat formulation on the second line, $g_{\mu\nu} = \Omega^2 \eta_{\mu\nu}$, where the \emph{tortoise coordinate} $x$ is defined through $dx = dr/f(r)$, and the \emph{conformal factor}  is $\Omega(x)=\sqrt{f(r(x))}$. $-\infty<x<\infty$ covers the region outside the horizon, $r > 2M$. The causal structure of this spacetime is essentially the same as the $3+1$-dimensional Schwarzschild black hole.

Metric~\eqref{eq:11metric} provides the $1+1$ decomposition
\begin{align}\label{eq:A_phi_tortoise}
    \phi = -\alpha^{-1} X_t = \Omega^{-1} X_t  \ \ ,  \ \ A_x= X_x \ .
\end{align}
The fact that $g_{\mu\nu}$, $\alpha$ and $\gamma_{xx}$ all vanish on the horizon due to $\Omega(-\infty)=0$ will be a central point in the subsequent discussion. 

We used the same methods as in \textcite{Coates:2022qia} to evolve NPT on the spacetime~\eqref{eq:11metric}, which were in turn adapted from \textcite{Clough:2022ygm}. We scale field values and coordinates to set $\mu^2=\lambda=1$ (we are only interested in $\lambda>0$), which means the only physically meaningful parameter is the dimensionless $M\mu$. Our results are for $M\mu=1$, but the outcome is qualitatively similar for any $M\mu$, for which only time and length scales change. In all cases, the only nonvanishing part of the initial data is a narrow Gaussian for $X_x(t=0,x)$, which satisfies the constraint equation.

\begin{figure}
\begin{center}
\includegraphics[width=.48\textwidth]{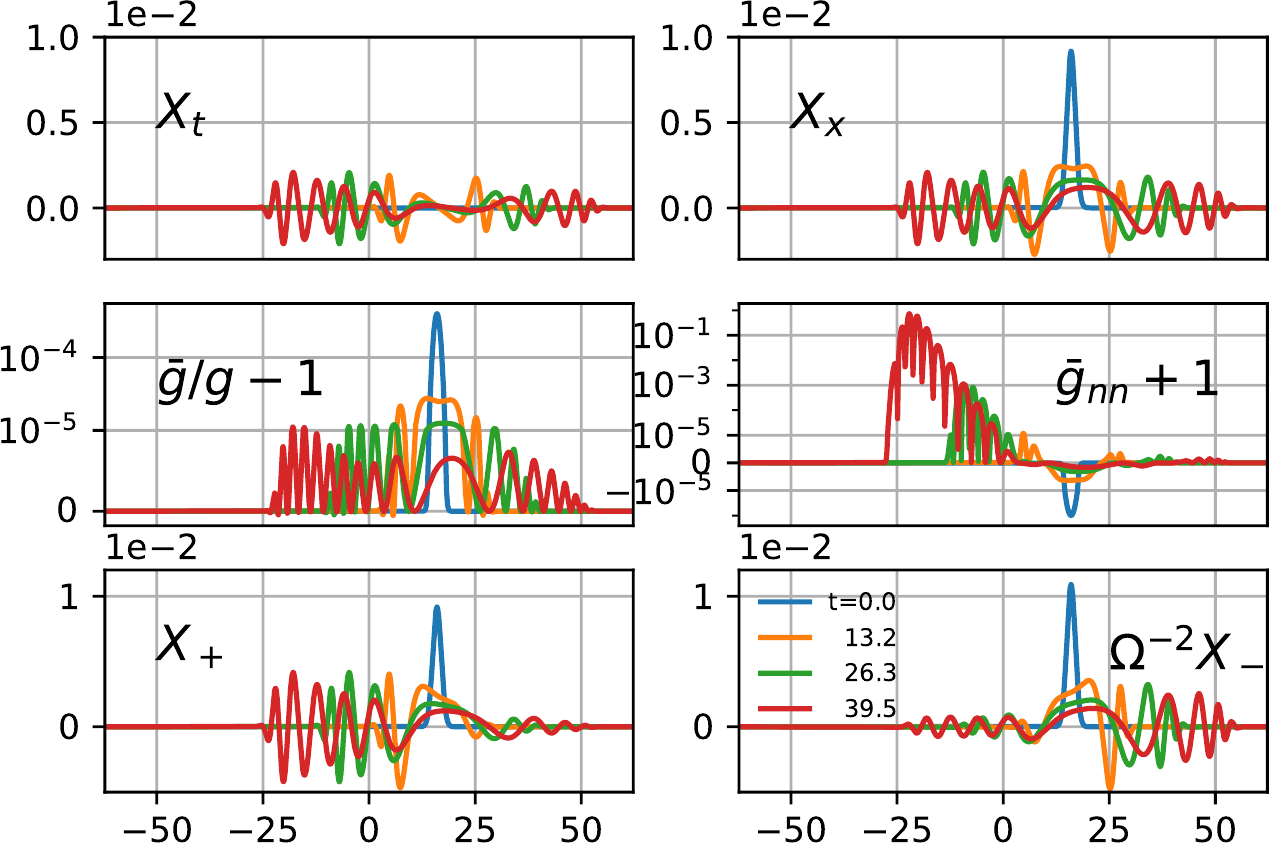}
\end{center}
\caption{Snapshots of $X_\mu$ and $\gbar_{\mu\nu}$ as a time symmetric  wave packet breaks up into two pieces, one falling into a black hole and the other moving out in tortoise coordinates (Eq.~\ref{eq:11metric}), $\mu^2=1, \lambda=1$. Our discussion concentrates on the ingoing piece (moving to the left) which eventually causes a coordinate singularity, $\gnn=0$. First row: The field components $X_t$, $X_x$ have steady amplitudes. Second row: $\gnn=0$ is achieved due to the amplifying effect of $g^{\mu\nu}\sim \Omega^{-2}$ on the steady values of the vector components. However, $\gbar/g \approx 1$ does not show any sign of physical breakdown, implying $|X^2|$ is not growing. Third row: Non-growth of $X^2= \Omega^{-2} X_+ X_-$ is revealed in the behavior of $X_\pm=X_t\pm X_x$.}
\label{fig:X_mu_gbar}
\end{figure}
The time evolution of an initially low amplitude wave packet in the spacetime of~\eqref{eq:11metric}  can be seen in Fig.~\ref{fig:X_mu_gbar}. In terms of the vector components $X_\mu$ the evolution looks mundane (the first row), the ingoing packet attains a constant amplitude for both $X_t$ and $X_x$ as it approaches the horizon. However, one can see a steady growth in $\gnn$, which eventually satisfies $\gnn=0$, breaking down the numerical evolution as we discussed before (second row). That is, the constraint criterion is satisfied in this example. 

Meanwhile, the effective metric never approaches a singularity in Fig.~\ref{fig:X_mu_gbar} as seen in its determinant $\gbar$, hence, the physical time evolution is completely healthy by the metric singularity criterion. Note that this is also the case in flat spacetime when the initial amplitude of the vector field is low enough~\cite{Coates:2022qia}, as we discussed in the beginning of this section. In light of these, our aim is showing that $\gnn=0$ is a coordinate effect by evolving the same system in other foliations where $\gnn=0$ is not encountered.

Let us first understand the singularity arising from $\gnn=0$ better before we see how we remove it. Note that the steady amplitude of the components of $X_\mu$ easily explains how $\gnn=0$ is reached. Since $\gnn= -1+ \Omega^{-2}\left[ -(X_x)^2 +3 (X_t)^2 \right]$, the eventual vanishing of $\gnn$ is guaranteed, since the $\Omega^{-2}$ factor arising from $g^{\mu\nu}$ grows without bound. Alternatively, in the $1+1$ formulation, $\gnn= -1+ \left[ -A_xA^x +3 \phi^2 \right]$, and $A^x = \gamma^{xx} A_x =\Omega^{-2} A_x$ and $\phi$ diverge due to the vanishing spatial metric and the shift, respectively [cf. Eq.~\eqref{eq:A_phi_tortoise}].

The second important observation is that $\gbar/g$ never deviates far from $1$, which shows the nonequivalance of the two criteria of breakdown. However, this is also mathmematically perplexing. Recall that $\gbar/g$ only depends on the norm of the vector field [cf. Eq.~\eqref{eq:detg}], and the fact that it changes insignificantly means $|X^2|=\Omega^{-2} | (X_x)^2 - (X_t)^2 | = |-A_xA^x + \phi^2|\ll 1$ throughout the evolution. However, by our previous argument about the behavior of the vector field components, unless there is a large cancellation in $(X_x)^2 - (X_t)^2 $, we would expect $|X^2|$ to grow as the horizon is approached due to the $\Omega^{-2}$ factor. A large cancellation indeed occurs due to the behavior of $X_\pm=X_x \pm X_t$, which can be seen in the last row of Fig.~\ref{fig:X_mu_gbar}, revealing that $X_-\sim \Omega^2$ in this region. Thus, $X^2=\Omega^{-2} X_+ X_-$ has asymptotically constant amplitude near the horizon. 

The behavior of $X_{t,x,\pm}$ can also be understood in surprising detail by an analytical study of the near-horizon behavior of the vector field, which shows that any small amplitude initial data leads to an ingoing wave with a roughly constant amplitude for $X_t$, $X_x$ and $X^2$, see the appendix. This also proves that $\gnn=0$ is encountered \emph{generically} in this scenario, which was further confirmed by numerical computation.

\begin{figure}
\begin{center}
\includegraphics[width=.48\textwidth]{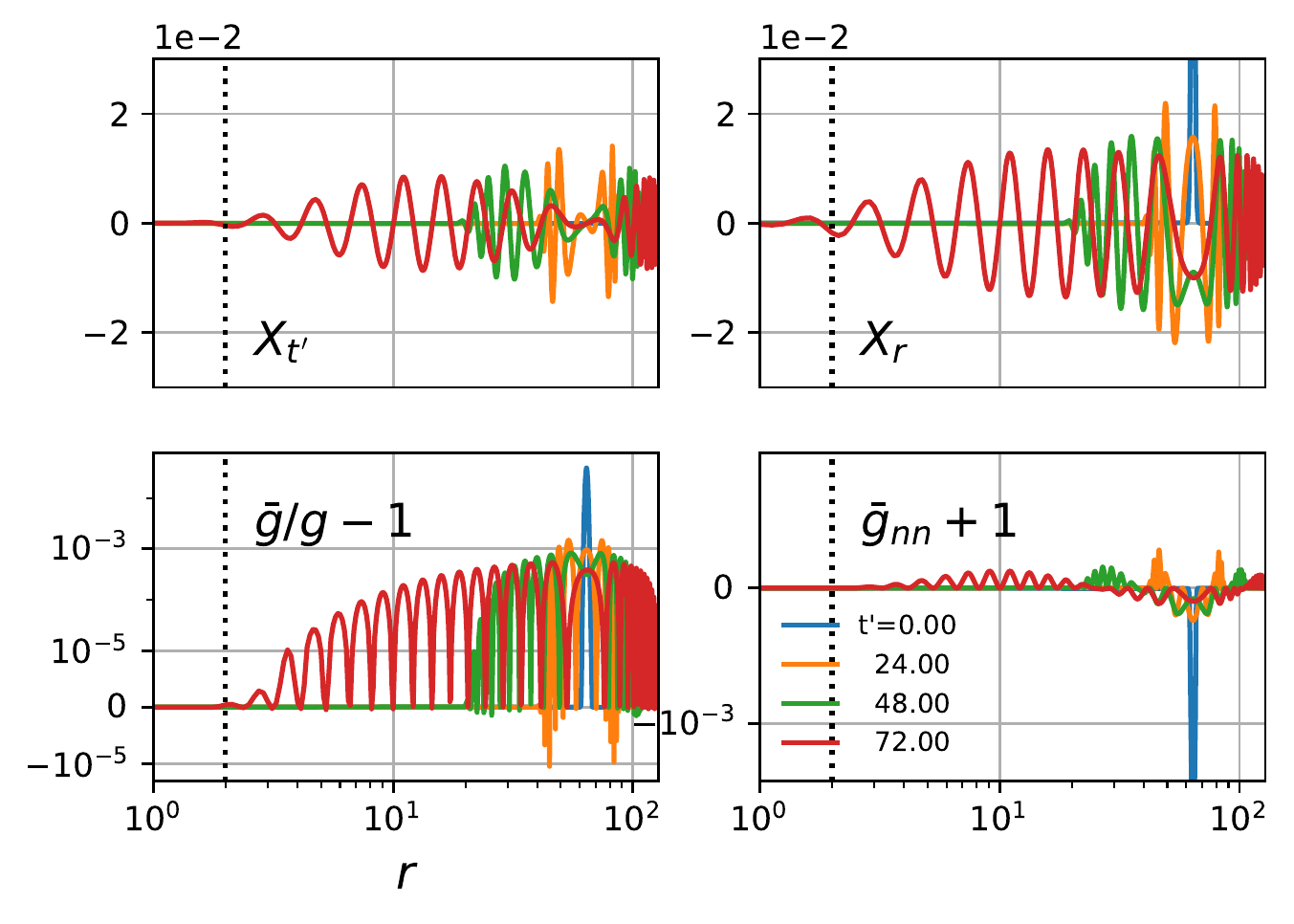}
\end{center}
\caption{
Snapshots of $X_\mu$ and $\gbar_{\mu\nu}$ as a wave packet falls into a black hole in the Cook-Scheel coordinates (Eq.~\ref{eq:cs_metric}), $\mu^2=1, \lambda=1$. The horizon is at $r=2M=2$, marked by the vertical dotted line. Similarly to Fig.~\ref{fig:X_mu_gbar}, the field components $X_{t',r}$ have steady amplitudes as they approach the horizon. However, $A^r$ and $\phi$ do not grow arbitrarily large thanks to nonvanishing lapse, shift and spatial metric, and $\gnn=0$ is not encountered anywhere. The initial amplitude of $X_r$ is higher than that of $X_x$ in Fig.~\ref{fig:X_mu_gbar} for this sample case.
}
\label{fig:CS}
\end{figure}
We will finally show that the constraint criterion is a coordinate-dependent one, and the problem it indicates can be removed by a change of coordinates. Let us consider the same spacetime in the Cook-Scheel (CS) coordinates~\cite{Cook:1997qc}
\begin{align}\label{eq:cs_metric}
    ds^2 = - F^{-2} dt'^2 + F^2 \left(dr+u^2F^{-2} dt' \right)^2 \
\end{align}
where $u=\sfrac{2M}{r}$ and $F^2=(1+u)(1+u^2)$. The horizon is located at $r=2M$.

CS coordinates have the \emph{horizon-penetrating} property where the spacetime metric as well as all the terms of the $d+1$ decomposition, such as $\alpha =F$, $\beta^r=u^2F^{-2}$, $\gamma^{rr}=F^{-2}$, are finite on and inside the horizon. Recall that the main problem of the tortoise coordinates in the $1+1$ decomposition was the unrestricted growth of $g^{\mu\nu}$, $\alpha^{-1}$ and $\gamma^{xx}$. Hence, horizon-penetration could prevent the growth of $\gnn$, and it is something we look for. On the other hand, the behaviour of the CS foliation in the asymptotically far region $r \to \infty$ is identical to that of the tortoise or Schwarzschild coordinates, so the growth problem is not merely moved to another part of the spacetime.

The evolution of an initially small Gaussian wave packet of $X_r$ using the CS coordinates and the associated foliation can be seen in Fig.~\ref{fig:CS}. The field components $X_{t',r}$ still have steady amplitudes as they approach the horizon. However, this time we expect $A^r$ and $\phi$ to be finite at the horizon. This is indeed the case. $\gnn=0$ is avoided altogether in contrast to Fig.~\ref{fig:X_mu_gbar}, and the evolution can continue until the ingoing packet reaches the physical singularity. Thus, the constraint criterion does not indicate any physical pathology in the time evolution of self-interacting vectors.

The original motivation for introducing the CS coordinates was satisfying the harmonic time slicing condition which has desirable properties for numerical relativity~\cite{Cook:1997qc}, but they turn out to be more appropriate for our purposes as well. We should emphasize that the CS coordinates are not unique in the above respect, coordinates that are finite and nonzero everywhere aside from the spacetime singularity, including asymptotic infinity, generally provide similar results.

\noindent {\bf \em  Discussion:}
Checking for the existence of problematic degrees of freedom, e.g. ghosts that grow exponentially, have been an integral part of model building. Indeed, many vector and tensor field theories have been ruled out this way, or their most extensive generalizations were constructed by following such guidelines~\cite{deRham:2014zqa,Heisenberg:2014rta}. The novelty of the two criteria we considered is that the theory does not carry such a problematic degree of freedom in all parts of the phase space, rather, the vector can evolve without issue for a finite time, and in some cases even indefinitely, but the time evolution can dynamically reach a point where it cannot be continued any more. This can be useful in guiding the efforts to explore the extensions of our current models in gravity, cosmology, high energy physics, and effective field theories in general, and examples are known in scalar-tensor theories~\cite{East:2022ppo,Corman:2022xqg,R:2022hlf}. However, it is important to ensure that the appearance of this dynamical breakdown is correctly identified so that accurate theoretical conclusions can be achieved.

One method for such identification has been checking whether the constraint equation of the theory can be unambiguously satisfied at all times, and we have shown that this criterion does not signify a physical breakdown. Rather, it reflects a shortcoming of the specific evolution scheme, Eq.~\eqref{eq:eom1p1}, for a specific coordinate choice. The problem completely disappears when we use more appropriate coordinates. The coordinate-dependent nature of the constraint criterion can be especially hard to notice if one works on a flat background spacetime as is usually the case in high energy theory~\cite{Mou:2022hqb}, where the idea of a foliation beyond the trivial one in Minkowski coordinates might even look unnatural and unnecessarily contrived. Though, our results make it clear that nontrivial foliations and curvilinear coordinates are essential for the study of self-interacting vectors.

The fact that the constraint criterion is equivalent to $\gnn=n^\mu n^\nu \gbar_{\mu\nu}=0$ already gives a hint about the foliation dependence, since the choice of $n^\mu$ defines the foliation. Furthermore, the constraint itself, $\mathcal{C}$, is an object that lives on the spatial hypersurfaces, hence the form of the equation ${\cal C}=0$ and whether it has a unique solution also depends on how we choose coordinates and foliate our spacetime. 

$d+1$ decomposition is ubiquitous in formulating the time evolution of tensor field theories in curved space, and highlights the constrained nature of the dynamics of the vector fields. This likely played a role in the misidentification of the constraint as the culprit of a physical breakdown. Nevertheless, we expect the results of the existing studies that used the constraint criterion to be essentially valid in that the time evolution indeed breaks down in their examples, even though it occurs at a different spacetime point which may or may not be covered in the computation.

We would want to emphasize that, even though the constraint criterion detects a coordinate singularity, self-interacting vector time evolution does physically break down as well, which can be identified using what we called the metric singularity criterion, $\gbar_{\mu\nu}$ becoming singular. It is also known that in certain parts of the $(\mu^2, \lambda)$ parameter space, one is guaranteed to encounter a coordinate singularity before the physical breakdown occurs if the evolution method of Eq.~\eqref{eq:eom1p1} is followed. Thus, it is crucial to use appropriate coordinates or novel time evolution techniques to circumvent such spurious problems~\cite{Coates:2022qia}. 

We removed the coordinate singularity by using an alternative coordinate system that is built in advance. However, more general and less symmetric evolutions likely require an adaptive foliation scheme which updates the normal vector $n^\mu$, hence the lapse and the shift, dynamically as the vector field evolves in time~\cite{Coates:2022qia}. Implementation of such schemes is a major future endeavor in the study of self-interacting vectors.

There are other avenues of exploration for the well-posedness of self-interacting vector field theories. The physical breakdown of the theory is a problem if we take it at face value, but it is sometimes possible to view action~\eqref{eq:action} as an \emph{effective field theory} which is to be completed in the ultraviolet section. Then, the problem is understanding how and if various issues we raised are resolved in the parent theory, which is actively studied~\cite{Aoki:2022woy,Barausse:2022rvg,Aoki:2022mdn}. On another avenue, it is still not known how backreaction affects the findings, since all results so far concentrated on fixed backgrounds.

\acknowledgements
We thank Will East for many enlightening discussions. F.M.R acknowledges support from T\"UB\.ITAK Project No. 122F097.

\appendix

\section{Understanding the near-horizon behavior through mode analysis}
The behavior of $X_\pm$ and $X^2$ in the tortoise coordinates can be understood  by solving the coupled equation system arising for them in the Proca limit $\lambda \to 0$. The solution is easiest using the densitized variables
\begin{align}\label{eq:calX_def}
    \cala_\pm =\Omega^{-1} X_\mp 
\end{align}
which have the advantage that their field equations are uncoupled
\begin{align}
    \left(-\partial_t^2 + \partial_x^2\right) \cala_\pm \mp 2 (\partial_x \ln \Omega) \partial_t \cala_\pm -{\cal V}(x) \cala_\pm &= 0 \ ,
\end{align}
where
\begin{align}
    {\cal V}(x) &= \mu^2 \Omega^2 + (\partial_x \ln \Omega)^2 - \partial_x^2 \ln \Omega
\end{align}

Proceeding with separation of variables $\cala_\pm = e^{i\omega t} \Xi_\pm(x)$, we obtain the eigenfunction problems
\begin{align}
    -\partial_x^2\ \Xi_\pm  + \left[ {\cal V}(x)\pm 2i\omega \frac{M}{r^2(x)} \right] \Xi_\pm &= \omega^2 \Xi_\pm\ .
\end{align}
In the near horizon region $x \to -\infty$, the leading terms reduce to
\begin{align}
    -\partial_x^2\ \Xi_\pm  &\approx \left[ \omega^2 \mp i \frac{\omega}{2M} - \frac{1}{16M^2} \right]\Xi_\pm \\
    &= \left[ \omega \mp i \frac{1}{4M} \right]^2 \Xi_\pm 
\end{align}
which means the ingoing modes behave as
\begin{align}
    \Xi_\pm (x)  &\approx e^{i\left( \omega \mp i \frac{1}{4M} \right) x} = e^{i\omega x} e^{\pm \frac{x}{4M}}
\end{align}
Noting the asymptotic behavior of the conformal factor $\Omega(x \to -\infty) \approx e^{x/4M}$,
\begin{align}\label{eq:calXpm_mode}
    \Xi_+(x \to -\infty) &\sim e^{ikx}\ \Omega\nonumber \\
    \Xi_-(x \to -\infty) &\sim e^{ikx}\ \Omega^{-1} \ ,
\end{align}
with $k=\omega$. Therefore, $X^2=\Omega^{-2}X_+X_-=\cala_+\cala_-$ has a steady amplitude near the horizon as seen in the numerical computations when $\lambda \neq 0$. A similar result can be obtained by analyzing the eigenfunctions for $\Omega^{\mp}  \cala_\pm$, which are plane waves $e^{ikx}$ near the horizon, an equivalent result to Eq.~\eqref{eq:calXpm_mode}.

To summarize, recovering $X_{t,x}$, the modes behave as
\begin{align}\label{eq:Xpm_mode}
    X_{t,x,+}(x \to -\infty) &\sim e^{i\omega t} e^{ikx}\ \nonumber \\
    X_-(x \to -\infty) &\sim e^{i\omega t} e^{ikx}\ \Omega^2(x)\ .
\end{align}
with $\omega=k$, which explains why ingoing $X_{t,x}$ wave packets retain their amplitude in Fig.~\ref{fig:X_mu_gbar}. Furthermore, $X_- \sim \Omega^{2}$, ensures that $|X^2|$ does not grow, hence there is no physical singularity [cf. Eq.~\eqref{eq:detg}].

\bibliography{references_all}

\end{document}